**Data-driven Exploration of Tropical Cyclone's Controllability**

Yohei Sawada[1], Masashi Minamide[1], Yuyue Yan[2,3], Kazumune Hashimoto[2], Le Duc[1]

[1] Department of Civil Engineering, Graduate School of Engineering, the University of Tokyo, Tokyo, Japan
[2] Division of Electrical, Electronic and Infocommunications Engineering, Graduate School of Engineering, the University of Osaka, Osaka, Japan
[3] now at Information Physics & Computing, Graduate School of Information Science and Technology, the University of Tokyo, Tokyo, Japan

Corresponding author: Y. Sawada, Department of Civil Engineering, the University of Tokyo, Tokyo, Japan, 7-3-1, Hongo, Bunkyo-ku, Tokyo, Japan, yoheisawada@g.ecc.u-tokyo.ac.jp

**Abstract**

Although the chaotic nature of the atmosphere may enable efficient control of tropical cyclones (TCs) via small-scale perturbations, few studies have proposed data-driven optimization methods to identify such perturbations. Here, we apply the recently proposed Ensemble Kalman Control (EnKC) to a TC simulation. We show that EnKC finds small-scale perturbations that mitigate TC. An EnKC-estimated reduction in surface water vapor, located approximately 250km from the TC center, suppresses convective activity and latent heat release in the eye wall, leading to a reduction of TC intensity. To advance the discovery of feasible TC mitigation strategies, we discuss the potential of this data-driven method for leveraging chaos, as well as its remaining challenges.

**Plain Language Summary**

Tropical cyclones are chaotic systems, which means that small changes to the atmosphere might change how strong they become. We tested an approach called Ensemble Kalman Control that searches for the effective "small tweaks" to weaken a storm in a realistic weather model. Our method found useful tweaks that act over small areas. The most effective change was to slightly reduce the amount of water vapor near the ocean surface about 250 kilometers from the storm's center. This reduces activities of the tall thunderstorms and the release of heat in the ring of tall clouds, which in turn lowered the storm's intensity. Our results suggest that targeted and data-driven interventions could someday help limit cyclone damage. However, they also highlight practical challenges.

**Key points**

We propose a novel data-driven, or top-down, method to identify perturbations to mitigate tropical cyclones.

Our method effectively leverages chaotic nature of the atmosphere to control the system.

# 1. Introduction

On 16 September 1961, an aircraft carrying eight canisters of silver iodide flew into the eyewall of Tropical Cyclone (TC) Esther to conduct cloud seeding. Another aircraft subsequently observed a reduction in kinetic energy near the eyewall after the seeding. This marked the beginning of mankind's attempts to mitigate TCs. Although the effect of cloud seeding on TC intensity was not confirmed as robust during Project STORMFURY (1962-1981) (Willoughby et al. 1985), many subsequent simulation-based studies have explored alternative TC mitigation methods, such as offshore wind turbines (Jacobson et al. 2014), power-generating sailing ships (Horinouchi and Mitsuyuki 2023), aerosol injection (Zhang et al. 2007; Cotton et al. 2007; Tran et al. 2025), and sea surface cooling to reduce evaporation (Latham et al. 2012; Hlywiak and Nolan 2022; Oceantherm 2025) (see Miller et al. (2023) for a comprehensive review).

It is very likely that the atmosphere is a chaotic system characterized by strong sensitivity to small perturbations. In such systems, small control interventions can dramatically alter future states, potentially enabling highly efficient control. This concept of chaos control has been investigated across many scientific disciplines (Shinbrot et al. 1993), and some meteorologists have noted its potential for weather modification (Hoffman, 2002; Henderson et al. 2005). However, most previous studies on the modifications of TC and other weather systems did not explicitly consider or leverage this sensitivity to small perturbations. Moreover, this same sensitivity limits predictability. Thus, the atmosphere presents a dual nature of both controllability and unpredictability, and it is generally difficult to control unpredictable systems. Previous studies have not fully addressed this dilemma, as they relied on "perfect model" experiments, in which one simulation with no control force was treated as a truth despite their longer simulation period than a predictability limit and compared with a controlled counterpart. In other words, previous works assessed the controllability of the atmosphere assuming infinite predictability. In these "perfect model" experiments, small-scale perturbations which leverage chaos cannot be robustly found.

Miyoshi and Sun (2022) proposed a useful framework to assess the controllability of the atmosphere by leveraging its chaotic nature. Their contributions were twofold. First, they introduced the Control Simulation Experiment (CSE) as an extension of the well-known Observing System Simulation Experiment (OSSE; Arnold and Dey 1986; Hoffman and Atlas 2016). In a CSE, a synthetic "nature run" is generated by simulation. Observations derived from this nature run are assimilated into a model estimate of the system state.

Then, based on this state estimate by data assimilation, a controller estimates appropriate interventions and adds them to the nature run. Although the nature run is generated by simulation, it is inaccessible to the controller, who must act under finite predictability imposed by data assimilation. CSE is useful to analyze the balance between controllability and predictability in chaotic systems (see also Miyoshi 2025). Second, Miyoshi and Sun (2022) proposed using ensemble forecasts to find small but effective control perturbations. From the analysis ensemble estimated by ensemble data assimilation, the extended ensemble forecast is performed. Then, they defined the "best" and "worst" ensemble members and used the difference between these two members to design control perturbations. They found that this approach effectively exploited the system's chaotic dynamics to estimate appropriate perturbations. Sawada (2024) advanced the ensemble-based method by proposing Ensemble Kalman Control (EnKC). Recognizing that data assimilation methods in geoscience such as Ensemble Kalman Filter (EnKF) and Model Predictive Control (MPC) in control engineering both minimize similar cost functions, Sawada (2024) proposed using EnKF to estimate control perturbations effectively, leveraging ensemble-based sensitivity information.

Inspired by Miyoshi and Sun (2022), many studies have performed CSE and explored the controllability of chaotic systems (Sun et al. 2023; Ouyang et al. 2023; Kawasaki and Kotsuki 2024; Sawada 2024, 2025; Kawasaki et al. 2025; Kurosawa et al. 2025). However, these works used toy models such as those of Lorenz (1963) and Lorenz et al. (1995). To our knowledge, no published studies have yet to apply CSE or ensemble forecast-based control optimization to realistic atmospheric simulation. Although EnKC was designed to apply high-dimensional geoscientific problems, its potential has not been examined in weather modification problems. Here, we present the first application of EnKC to a TC simulation. Our goal is not to demonstrate that TCs are controllable, but rather to propose a novel and useful data-driven framework to explore when, where and how to intervene in TCs to mitigate their intensity by leveraging their inherent chaotic nature.

## 2. Method
### 2.1. Ensemble Kalman Control (EnKC)
The Ensemble Kalman Filter (EnKF; Evensen 1994) is a widely used data assimilation method in which observations and numerical simulations are integrated to provide accurate state estimates. EnKF minimizes the following cost function:

$$J(x_t) = \frac{1}{2}\left(x_t - \overline{x_t^b}\right)^T {P^b}^{-1}\left(x_t - \overline{x_t^b}\right) + \frac{1}{2}\left(y_t^o - H(x_t)\right)^T R^{-1}\left(y_t^o - H(x_t)\right) \quad (1)$$

where $x_t$ is the state variables at time $t$, $\overline{x_t^b}$ is the background ensemble mean of state estimates, $P^b$ is the background error covariance matrix estimated from ensemble, $y_t^o$ is the observation at time $t$, $H$ is the observation operator, and $R$ is the observation error covariance matrix. There are many flavors of EnKF to obtain the solution of the minimization of Equation (1) and generate the analysis ensemble members, $x_t^{a(i)}$ ($i = 1,2,3,…,N$, where $N$ is the ensemble size). See Houtekamer and Zhang (2016) for the comprehensive review of EnKF.

In EnKC, the minimization of Equation (1) is first performed using observations to obtain the analysis ensemble, $x_t^{a(i)}$. Then, an extended ensemble forecast is performed from $x_t^{a(i)}$ over a prediction horizon, $T_c$. By performing Ensemble Kalman Smoother (EnKS; Evensen and van Leeuwen 2000), the following minimization problem is solved to estimate an appropriate control perturbation:

$$J^c(x_t) = \frac{1}{2}(x_t - \overline{x_t^a})^T P^{a-1}(x_t - \overline{x_t^a}) + \frac{1}{2}(r_{t+T_c} - H^c(x_{t+T_c}))^T R_c^{-1}(r_{t+T_c} - H^c(x_{t+T_c})) \quad (2)$$

$$s.t. \, x_{t+k+1} = M(x_{t+k}), k = 0,1,…,T_c - 1$$

where $\overline{x_t^a}$ is the analysis ensemble mean, $P^a$ is the analysis error covariance matrix, $r_t$ is the control target indicating the desired system state at time $t$, $H^c$ is the operator that projects the state variables to the control target, enabling comparison between the forecasted state and the control target, $R_c$ is the user-defined weights, and $M$ is the model which describes the evolution of the states. This minimization seeks the smallest perturbation, $x_t - \overline{x_t^a}$, to be added to the initial state of nature to effectively reduce the difference between the control target and the forecasted future state. Because this process is conceptually similar to MPC, EnKC can be recognized as a type of MPC in which the minimization of the cost function is solved by EnKS (see also Kurosawa et al. 2025). In EnKC, control targets are treated as pseudo-observations, and optimal perturbations are obtained by "assimilating" these control targets into the model-estimated state variables. Further algorithmic details are provided in Sawada (2024).

The solution of the vanilla EnKC, $x_t^c - \overline{x_t^a}$, includes non-zero elements for all state variables, implying modifications to the entire system state. This is apparently unrealistic in the case of TC modification. To enforce sparsity in control interventions, we applied the following function to all entries of $x_t^c - \overline{x_t^a}$:

$$T(\theta_i) = \begin{cases} 0 & if \, \frac{|\theta_i|}{\sigma_i} < \Lambda \\ \theta_i & otherwise \end{cases} \quad (3)$$

$$\Lambda = \lambda * \max\left(\frac{|\theta_i|}{\sigma_i}\right)$$

where $\theta_i$ is the ith entry of $x_t^c - \overline{x_t^a}$ and $\sigma_i$ is the analysis standard deviation of the ith entry of a state vector. The ratio $\frac{|\theta_i|}{\sigma_i}$ can be interpreted as a signal-to-noise ratio of the control perturbations, and we used only perturbations with sufficiently large signal-to-noise ratios. The hyperparameter $\lambda$ determines sparsity. When $\lambda = 1$, only the grid point with the largest signal-to-noise ratio is perturbed. Smaller $\lambda$ allow interventions across more state variables, and when $\lambda = 0$, Equation (3) reduces to the vanilla EnKC.

## 2.2. Experiment design

We used SCALE-LETKF (Lian et al. 2017), which couples the Scalable Computing for Advanced Library and Environment (SCALE) regional atmospheric model (Nishizawa et al. 2015; Sato et al. 2015) with the Local Ensemble Transform Kalman Filter (LETKF) data assimilation system (Hunt et al. 2007; Miyoshi and Yamane 2007). The SCALE-LETKF system has been successfully applied to forecasting various severe weather events (e.g., Honda et al. 2018a, 2018b, 2025; Taylor et al. 2021a, 2023).

We simulated an idealized TC within a 2000km × 2000km horizontal domain with a horizontal grid spacing of 5km. The model employed 50 vertical levels with the model top at 25km. The initial environment was horizontally homogeneous and defined by the mean sounding profile of Jordan (1958). The initial vortex had a maximum wind speed of 20m/s and a radius of maximum wind of 120km. The Coriolis parameter was set to $5 \times 10^{-5}$ [1/s]. Periodic boundary conditions were applied. We used the same setting of physical parameterization as previous SCALE-LETKF studies (e.g., Honda et al. 2025). The model employed a Smagorinsky-type turbulence parameterization (Brown et al. 1994), the Mellor-Yamada-Nakanishi-Niino boundaly layer scheme (Nakanishi and Niino 2004), a parallel plane radiation model (Sekiguchi and Nakajima 2008), and a one-moment six-category bulk cloud microphysics model (Tomita 2008). Sea surface temperature was set to 300 K, and surface fluxes were estimated by a bulk model.

We performed a CSE experiment. We added Gaussian white noise whose mean and standard deviation are 0 and 0.1 g/kg to the initial vertical profile of water vapor and generated 101 initial conditions. One of these initial conditions was randomly selected as the "nature run", while the remaining 100 initial conditions were used for the ensemble members of EnKC, with the ensemble size of 100. The same SCALE model was used to integrate both the nature run and the ensemble members of EnKC. After a 72-hour spin-up period, we performed data assimilation for 96 hours. Subsequently, the intervention phase was initiated and continued for 72 hours during which the TC was in its stable stage.

We assumed that wind speed, specific humidity, and temperature can be observed in every 2 grid point horizontally and vertically, with the observation error of 0.1m/s, 0.01 g/kg, and 0.1 K, respectively. Also, the central pressure was assumed to be observed with an observation error of 5 hPa. Observations were generated by adding Gaussian white noise with a mean of 0 and a standard deviation equal to the assumed observation error to the nature run. We assimilated these observations every 1 hour. The horizontal and vertical localization radii were set to 50km and 0.3$ln$ p (where p is the pressure), respectively. A multiplicative covariance inflation factor of 1.45 was applied. Although this observation network allows accurate estimation of TC states, our controller had no access to nature run and needed to estimate perturbations under finite predictability.

The control target was defined as achieving a minimum pressure of 960 hPa at the lowest atmospheric layer (note that this is not sea-level pressure). In Equation (2), $r_{t+T_c}$ was set to 960 hPa, and the operator $H^c$ extracted the minimum pressure at the lowest layer from each ensemble member. To minimize the cost function (Equation 2), this central pressure value of 960 hPa was "assimilated" into $x_t^{a(i)}$. The prediction horizon, $T_c$, was set to 1 hour, so that control perturbations were estimated and applied every 1 hour based on a 1-hour extended forecast. Controlled nature was integrated by SCALE after EnKC-estimated perturbations had been added. The weighting parameter $R_c$ was set to 1.0 hPa. Since the central pressure at the beginning of the intervention period was approximately 955 hPa, our controller estimated perturbations to increase the central pressure.

In this study, we examined where water vapor should be removed by e.g., sea surface temperature cooling (Hlywiak and Nolan 2022; Oceantherm 2025), surfactant (e.g., Mozafari et al. 2019; Saggari and Bachi 2018; Gallego-Elvira et al. 2013; Schouten et al. 2012), or atmospheric water harvesting (e.g., Lord et al. 2021) to mitigate TC. Since removing water vapor at high altitudes is unlikely to be technologically feasible, modifications were restricted to the lowest atmospheric layer. In addition, interventions were limited to within a 500km radius from the TC eye. Therefore, all entries of $x_t^c - \overline{x_t^a}$ were set to zero except for the water vapor variables in the lowest layer near the TC center. Then, we applied the thresholding function of Equation (3) to further limit the perturbed model grid points. Finally, we ignored the positive water vapor perturbations, as increasing water vapor is also technologically unrealistic. We conducted experiments with $\lambda = 0.0, 0.5, 0.8, 0.9, 0.925$, and $0.95$ to examine how small the scale of interventions could be while still mitigating the TC.

## 3. Results

Figure 1 shows the control perturbations of water vapor added to nature run. The intervention begins northwest of the TC center and gradually moves counterclockwise with all $\lambda$ (see also Supplement Movies S1-S6 in https://drive.google.com/drive/folders/1rUxVDR2TC0dQQnScZLT8-NUv_wIUuZbL?usp=sharing). The control perturbations tend to appear on the side of active convection (indicated by black dots in Figure 1), suggesting that EnKC preferentially targets regions of strong moist convection. As $\lambda$ increases, the spatial scale of the interventions decreases. This effect is clearly illustrated in Figure 2, which shows the total amount of water vapor reductions, and Figure S1, which shows the total number of interventions. In the experiment of $\lambda = 0$, nearly all grid points in the lowest atmospheric layer around the TC center were modified, and some grid points were repeatedly perturbed throughout the 72-hour intervention period (note that we have 72 chances to intervene in our 72-hour intervention period). In contrast, experiments with larger $\lambda$ show interventions confined to smaller areas. To mitigate TC, water vapor reductions are applied approximately 250km from the TC center, surrounding the strongly convective eyewall region. The total amount of water vapor reduction is apparently smaller with the larger $\lambda$, so that the increase of $\lambda$ can mitigate the total energy necessary to intervene as well as the spatial scale of the interventions.

This intervention to water vapor successfully decreases TC intensity in most cases. Figure 3a shows that minimum sea-level pressure was increased by interventions in all experiments except for $\lambda = 0.95$. While the TC starts weakening in the first 24 hours in the experiments with smaller $\lambda = 0, 0.5, 0.8$, it takes longer to find the distinct effect of interventions in those with larger $\lambda$. Note that EnKC reduces the magnitude of perturbations when the central pressure approaches the prescribed control goal. Although the perturbations estimated with the experiments with smaller $\lambda$ can further increase TC pressure, EnKC does not exploit this potential once their control goals are nearly met. Ultimately, similar levels of TC weakening were achieved for most $\lambda$ values except for $\lambda = 0.95$, demonstrating that EnKC can efficiently identify perturbations that mitigate TC intensity. Although the relationship between central pressure and wind speed is not strictly linear, Figure 3b shows that these increases of minimum sea-level pressure were accompanied by substantial reductions in maximum surface wind speed.

The mitigation of TC intensity by localized water vapor removal can be attributed to the suppression of convective activity and the associated decrease in latent heat release within the eyewall. Figure 4 and Figure S2 reveal that all successful experiments produce substantial reductions in condensed water around the radius of maximum wind and weakened secondary flow by t = 48 h. This weakened convective activity cools the warm core and reduces the primary circulation and storm intensity, consistent with explanations based on the Sawyer-Eliassen equation (e.g., Pendergrass and Willoughby 2009). It should be noted that experiments with small-scale interventions (e.g., $\lambda = 0.9$) achieve comparable reductions in condensed water to those with larger-scale interventions (e.g., $\lambda = 0.0, 0.5$), implying that EnKC can pinpoint effective locations and timings for efficiently weakening convection. The regions of intervention in the experiments with larger $\lambda$ imply that moisture reduction upstream of the convectively active region may have contributed to the convective suppression. EnKC thus leverages small-scale perturbations of water vapor to affect strong convection in the eyewall and alter TC structure without requiring explicit prior knowledge of TC dynamics. In the case of failure (i.e., $\lambda = 0.95$), no consistent reduction in condensed water within the eyewall was found. Because our approach is intrinsically probabilistic, the overall success rate of interventions should be assessed in future works to confirm their robustness towards real-world applications, which is beyond the scope of this study.

## 4. Discussions and conclusions

Although we do not propose operationally feasible TC intervention methods in this paper, the spatial scale of our obtained interventions is smaller than that of many previous works. Earlier studies on TC modification typically considered large-scale interventions encompassing the entire storm (e.g., Henderson et al. 2005; Jacobson et al. 2014; Zhang et al. 2007; Cotton et al. 2007; Tran et al. 2025). For example, Hlywiak and Nolan (2022) examined the impact of targeted artificial ocean cooling to reduce evaporation (see also Oceantherm 2025) on the mitigation of TC and concluded that such interventions would be infeasible, as they would require an intervened area on the order of $10^5$ km$^2$. Our EnKC results suggest that the scales of intervention at each time can potentially be smaller than $10^3$ km$^2$, if the intervention location is adaptively updated over time, even under finite predictability, in which the controller relies only on imperfect 1-hour forecasts. However, our interventions' magnitude, which is approximately 0.5 g/kg reduction in near-surface (500m) water vapor, is still substantial, corresponding to roughly a 25-50% reduction in surface evaporation. While challenging, such magnitudes may not be impossible since surfactant intervention is reported to be able to reduce evaporative flux by up to 50% under idealized conditions (e.g., Mozafari et al. 2019). Nevertheless, their performance in the real ocean and the potential environmental side effects should be investigated. Also, the adaptive change of the intervention locations is operationally difficult. Although the present results do not confirm the feasibility of practical TC modification, the efficiency of the control intervention can be further improved through higher spatial resolution of atmospheric models to capture smaller-scale processes, the combinations of multiple intervention strategies, fine-tuning of hyperparameters such as prediction horizon ($T_c$) and control weights ($\boldsymbol{R}_c$), and the refinement of the algorithm to enhance robustness to non-linear dynamics and to explicitly consider realistic control constraints such as the distance of locations between subsequent interventions. This paper is an initial step to find effective and feasible perturbations to mitigate TCs by leveraging their chaotic nature under finite predictability.

Previously, numerical studies on TC and general weather modifications have adopted a process-driven, or bottom-up, approach. Based on physical understanding of targeted weather systems, researchers pre-determined when, where, and how to apply interventions, and then evaluated their effects by comparing simulations with and without prescribed controls. This research process is time-consuming for trials and errors and is difficult to leverage the chaotic nature of the atmosphere, since small changes in the location or magnitude of interventions can drastically alter the outcome. In contrast, we produce a data-driven, or top-down, approach for weather modification. Given a specified control objective, our algorithm automatically determines when, where, and how interventions should be applied. The resulting interventions are derived entirely from (simulated) data yet remain physically interpretable within the context of TC dynamics. We can leverage strong sensitivity to perturbations (i.e., chaos) by adaptively using ensemble sensitivity quantified from short-term ensemble forecast where predictability holds to some extent. EnKC is a useful tool for doing TC modification research in a data-driven, or top-down, way. This is the dawn of the data-driven exploration of TC's controllability. Combined with the process-based approaches, our approach has the potential to accelerate the exploration of controllability of TC and the other weather phenomena.

**Acknowledgements**

This work was supported by the JST Moonshot R&D program (Grant JMPJMS2281). We thank Seiya Nishizawa and Ryuji Yoshida for their help to perform an idealized TC simulation.

**References**

Arnold, C. P., and C. H. Dey (1986), Observing-systems simulation experiments: Past, present, and future, Bulletin of the American Meteorological Society, 67(6), 687–695, https://doi.org/10.1175/1520-0477(1986)067<0687:OSSEPP>2.0.CO;2.

Brown, A. R., S. H. Derbyshire, and P. J. Mason (1994), Large-eddy simulation of stable atmospheric boundary layers with a revised stochastic subgrid model, Quarterly Journal of the Royal Meteorological Society, 120, 1485–1512.

Cotton, W. R., H. Zhang, G. M. McFarquhar, and S. M. Saleeby (2007), Should we consider polluting hurricanes to reduce their intensity, Journal of Weather Modification, 39, 70–73, https://doi.org/10.54782/jwm.v39i1.204.

Evensen, G. (1994), Sequential data assimilation with a nonlinear quasi-geostrophic model using Monte Carlo methods to forecast error statistics, Journal of Geophysical Research: Oceans, 99(C5), 10,143–10,162, https://doi.org/10.1029/94JC00572.

Evensen, G., and P. J. van Leeuwen (2000), An ensemble Kalman smoother for nonlinear dynamics, Monthly Weather Review, 128, 1852–1867, https://doi.org/10.1175/1520-0493(2000)128<1852:AEKSFN>2.0.CO;2.

Gallego-Elvira, B., V. Martínez-Alvarez, P. Pittaway, et al. (2013), Impact of micrometeorological conditions on the efficiency of artificial monolayers in reducing evaporation, Water Resources Management, 27, 2251–2266, https://doi.org/10.1007/s11269-013-0286-3.

Henderson, J. M., R. N. Hoffman, S. M. Leidner, T. Nehrkorn, and C. Grassotti (2005), A 4D-Var study on the potential of weather control and exigent weather forecasting, Quarterly Journal of the Royal Meteorological Society, 131, 3037–3051, https://doi.org/10.1256/qj.05.72.

Hlywiak, J., and D. S. Nolan (2022), Targeted artificial ocean cooling to weaken tropical cyclones would be futile, Communications Earth & Environment, 3, 185, https://doi.org/10.1038/s43247-022-00519-1.

Hoffman, R. N. (2002), Controlling the global weather, Bulletin of the American Meteorological Society, 83, 241–248, https://doi.org/10.1175/1520-0477(2002)083<0241:CTGW>2.3.CO;2.

Hoffman, R. N., and R. Atlas (2016), Future observing system simulation experiments, Bulletin of the American Meteorological Society, 97(9), 1601–1616, https://doi.org/10.1175/BAMS-D-15-00200.1.

Honda, T., and Coauthors (2018a), Assimilating all-sky Himawari-8 satellite infrared radiances: A case of Typhoon Soudelor (2015), Monthly Weather Review, 146, 213–229, https://doi.org/10.1175/MWR-D-16-0357.1.

Honda, T., S. Kotsuki, G. Y. Lien, Y. Maejima, K. Okamoto, and T. Miyoshi (2018b), Assimilation of Himawari-8 all-sky radiances every 10 minutes: Impact on precipitation and flood risk prediction, Journal of Geophysical Research: Atmospheres, 123, 965–976, https://doi.org/10.1002/2017JD027096.

Honda, T. (2025), Exploring the intrinsic predictability limit of a localized convective rainfall event near Tokyo, Japan, using a high-resolution EnKF system, Journal of the Atmospheric Sciences, 82, 177–195, https://doi.org/10.1175/JAS-D-24-0022.1.

Horinouchi, T., and T. Mitsuyuki (2023), Gross assessment of the dynamical impact of numerous power-generating sailing ships on the atmosphere and evaluation of the impact on tropical cyclones, Scientific Online Letters on the Atmosphere (SOLA), 19, 2023-008, https://doi.org/10.2151/sola.2023-008.

Houtekamer, P. L., and F. Zhang (2016), Review of the ensemble Kalman filter for atmospheric data assimilation, Monthly Weather Review, 144, 4489–4532, https://doi.org/10.1175/MWR-D-15-0440.1.

Hunt, B. R., E. J. Kostelich, and I. Szunyogh (2007), Efficient data assimilation for spatiotemporal chaos: A local ensemble transform Kalman filter, Physica D: Nonlinear Phenomena, 230(1–2), 112–126, https://doi.org/10.1016/j.physd.2006.11.008.

Jacobson, M., C. Archer, and W. Kempton (2014), Taming hurricanes with arrays of offshore wind turbines, Nature Climate Change, 4, 195–200, https://doi.org/10.1038/nclimate2120.

Jordan, C. L. (1958), Mean soundings for the West Indies area, Journal of the Atmospheric Sciences, 15, 91–97, https://doi.org/10.1175/1520-0469(1958)015<0091:MSFTWI>2.0.CO;2.

Kawasaki, F., and S. Kotsuki (2024), Leading the Lorenz-63 system toward the prescribed regime by model predictive control coupled with data assimilation, Nonlinear Processes in Geophysics, 31, 319–333, https://doi.org/10.5194/npg-31-319-2024.

Kawasaki, F., A. Okazaki, K. Kurosawa, T. Tsuyuki, and S. Kotsuki (2025), Model predictive control with foreseeing horizon designed to mitigate extreme events in chaotic dynamical systems, EGUsphere, Preprint, https://doi.org/10.5194/egusphere-2025-1785.

Kurosawa, K., A. Okazaki, F. Kawasaki, and S. Kotsuki (2025), Ensemble-based model predictive control using data assimilation techniques, Nonlinear Processes in Geophysics, 32, 293–307, https://doi.org/10.5194/npg-32-293-2025.

Latham, J., B. Parkes, A. Gadian, and S. Salter (2012), Weakening of hurricanes via marine cloud brightening (MCB), Atmospheric Science Letters, 13, 231–237, https://doi.org/10.1002/asl.402.

Lien, G. Y., T. Miyoshi, S. Nishizawa, R. Yoshida, H. Yashiro, S. A. Adachi, et al. (2017), The near-real-time SCALE-LETKF system: A case of the September 2015 Kanto-Tohoku heavy rainfall, Scientific Online Letters on the Atmosphere (SOLA), 13(1), 1–6, https://doi.org/10.2151/sola.2017-001.

Lorenz, E. N. (1963), Deterministic nonperiodic flow, Journal of the Atmospheric Sciences, 20, 130–141, https://doi.org/10.1175/1520-0469(1963)020<0130:DNF>2.0.CO;2.

Lorenz, E. N. (1995), Predictability – a problem partly solved, in Seminar on Predictability, ECMWF, https://www.ecmwf.int/node/10829.

Lord, J., A. Thomas, N. Treat, et al. (2021), Global potential for harvesting drinking water from air using solar energy, Nature, 598, 611–617, https://doi.org/10.1038/s41586-021-03900-w.

Miller, J., A. Tang, T. L. Tran, R. Prinsley, and M. Howden (2023), The feasibility and governance of cyclone interventions, Climate Risk Management, 41, 100535, https://doi.org/10.1016/j.crm.2023.100535.

Miyoshi, T., and S. Yamane, (2007), Local Ensemble Transform Kalman Filtering with an AGCM at a T159/L48 Resolution. Monthly Weather Review., 135, 3841–3861, https://doi.org/10.1175/2007MWR1873.1.

Miyoshi, T., and Q. Sun (2022), Control simulation experiment with Lorenz's butterfly attractor, Nonlinear Processes in Geophysics, 29, 133–139, https://doi.org/10.5194/npg-29-133-2022.

Miyoshi, T. (2025), A duality principle for chaotic systems: From data assimilation to efficient control, Preprint, Research Square, https://doi.org/10.21203/rs.3.rs-7528999/v1.

Mozafari, A., B. Mansouri, and S. F. Chini (2019), Effect of wind flow and solar radiation on functionality of water evaporation suppression monolayers, Water Resources Management, 33, 3513–3522, https://doi.org/10.1007/s11269-019-02313-9.

Nakanishi, M., and H. Niino (2004), An improved Mellor–Yamada level-3 model with condensation physics: Its design and verification, Boundary-Layer Meteorology, 112, 1–31, https://doi.org/10.1023/B:BOUN.0000020164.04146.98.

Nishizawa, S., H. Yashiro, Y. Sato, Y. Miyamoto, and H. Tomita (2015), Influence of grid aspect ratio on planetary boundary layer turbulence in large-eddy simulations, Geoscientific Model Development, 8(10), 3393–3419, https://doi.org/10.5194/gmd-8-3393-2015.

Pendergrass, A. G., and H. E. Willoughby (2009), Diabatically Induced Secondary Flows in Tropical Cyclones. Part I: Quasi-Steady Forcing. Monthly Wether. Review, 137, 805–821, https://doi.org/10.1175/2008MWR2657.1.

Oceantherm (2025), Oceantherm, Website, https://www.oceantherm.no/ (accessed 2025-10-14).

Ouyang, M., K. Tokuda, and S. Kotsuki (2023), Reducing manipulations in a control simulation experiment based on instability vectors with the Lorenz-63 model, Nonlinear Processes in Geophysics, 30, 183–193, https://doi.org/10.5194/npg-30-183-2023.

Saggaï, S., and O. E. K. Bachi (2018), Evaporation reduction from water reservoirs in arid lands using monolayers: Algerian experience, Water Resources, 45, 280–288, https://doi.org/10.1134/S009780781802015X.

Sato, Y., S. Nishizawa, H. Yashiro, Y. Miyamoto, Y. Kajikawa, and H. Tomita (2015), Impacts of cloud microphysics on trade wind cumulus: Which cloud microphysics processes contribute to the diversity in a large-eddy simulation?, Progress in Earth and Planetary Science, 2(1), 23, https://doi.org/10.1186/s40645-015-0053-6.

Sawada, Y. (2024), Ensemble Kalman filter meets model predictive control in chaotic systems, Scientific Online Letters on the Atmosphere (SOLA), 20, 400–407, https://doi.org/10.2151/sola.2024-053.

Sawada, Y. (2025), Quest for an efficient mathematical and computational method to explore optimal extreme weather modification, Preprint, arXiv:2405.08387, https://arxiv.org/abs/2405.08387.

Schouten, P., and coauthors (2012), Evaluation of an evaporation suppressing monolayer system in a controlled wave tank environment: A pilot investigation, Australasian Journal of Water Resources, 16(1), 49–64, https://doi.org/10.7158/13241583.2012.11465403.

Sekiguchi, M., and T. Nakajima (2008), A k-distribution-based radiation code and its computational optimization for an atmospheric general circulation model, Journal of

Quantitative Spectroscopy and Radiative Transfer, 109, 2779–2793, https://doi.org/10.1016/j.jqsrt.2008.07.013

Shinbrot, T., C. Grebogi, J. Yorke, et al. (1993), Using small perturbations to control chaos, Nature, 363, 411–417, https://doi.org/10.1038/363411a0.

Sun, Q., T. Miyoshi, and S. Richard (2023), Control simulation experiments of extreme events with the Lorenz-96 model, Nonlinear Processes in Geophysics, 30, 117–128, https://doi.org/10.5194/npg-30-117-2023.

Taylor, J., T. Honda, A. Amemiya, S. Otsuka, Y. Maejima, and T. Miyoshi (2023), Sensitivity to localization radii for an ensemble filter numerical weather prediction system with 30-second update, Weather and Forecasting, 38, 611–632, https://doi.org/10.1175/WAF-D-21-0177.1.

Taylor, J., and Coauthors (2021), Oversampling reflectivity observations from a geostationary precipitation radar satellite: Impact on typhoon forecasts within a perfect model OSSE framework, Journal of Advances in Modeling Earth Systems, 13, e2020MS002332, https://doi.org/10.1029/2020MS002332.

Tomita, H. (2008), New microphysical schemes with five and six categories by diagnostic generation of cloud ice, Journal of the Meteorological Society of Japan, 86A, 121–142, https://doi.org/10.2151/jmsj.86A.121.

Tran, T. L., J. Fan, D. Rosenfeld, Y. Zhang, H. Cleugh, A. M. Hogg, and R. Prinsley (2025), Investigation of the sensitivity of tropical cyclogenesis to aerosol intervention, Journal of Geophysical Research: Atmospheres, 130, e2024JD041600, https://doi.org/10.1029/2024JD041600.

Willoughby, H., D. Jorgensen, R. Black, and S. Rosenthal (1985), Project STORMFURY: A scientific chronicle 1962–1983, Bulletin of the American Meteorological Society, 66(5), 505–514, https://doi.org/10.1175/1520-0477(1985)066<0505:PSASC>2.0.CO;2.

Zhang, H., G. M. McFarquhar, S. M. Saleeby, and W. R. Cotton (2007), Impacts of Saharan dust as CCN on the evolution of an idealized tropical cyclone, Geophysical Research Letters, 34, L14812, https://doi.org/10.1029/2007GL029876.

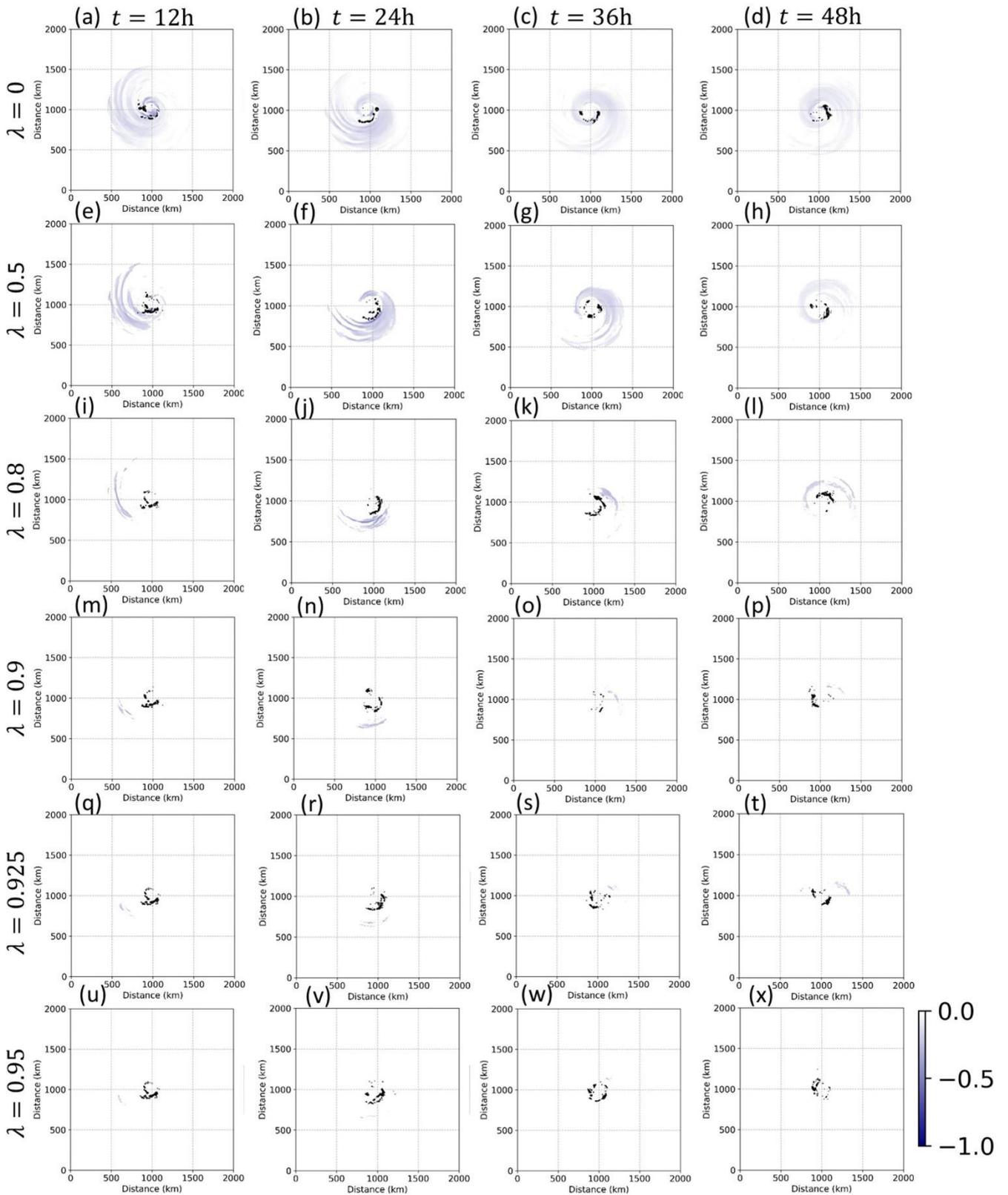

**Figure 1.** Control perturbation of water vapor [g/kg] at the lowest atmospheric layer estimated by EnKC with $\lambda$ of (a-d) 0.0, (e-h) 0.5, (i-l) 0.8, (m-p) 0.9, (q-t) 0.925, and (u-x) 0.95 at the time of (a,e,I,m,q,u) 12h, (b,f,j,n,r,v) 24h, (c,g,k,o,s,w) 36h, and (d,h,l,p,t,x) 48h after the beginning of the intervention. Black dots show the areas with vertical wind speed larger than 0.5 [m] at 7500m height.

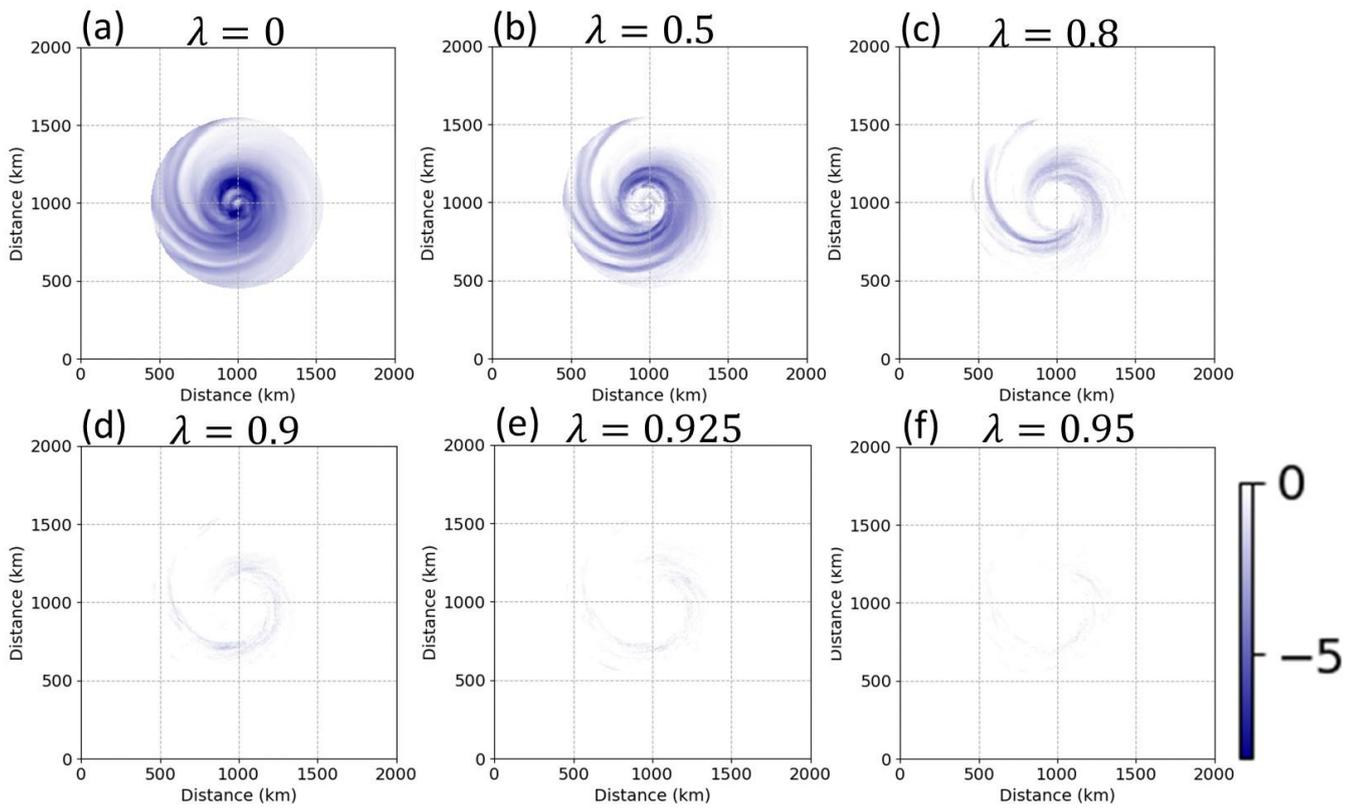

**Figure 2.** The total amount of water vapor changes [g/kg] at the lowest atmospheric level by EnKC interventions with $\lambda$ of (a) 0.0, (b) 0.5, (c) 0.8, (d) 0.9, (e) 0.925, and (f) 0.95.

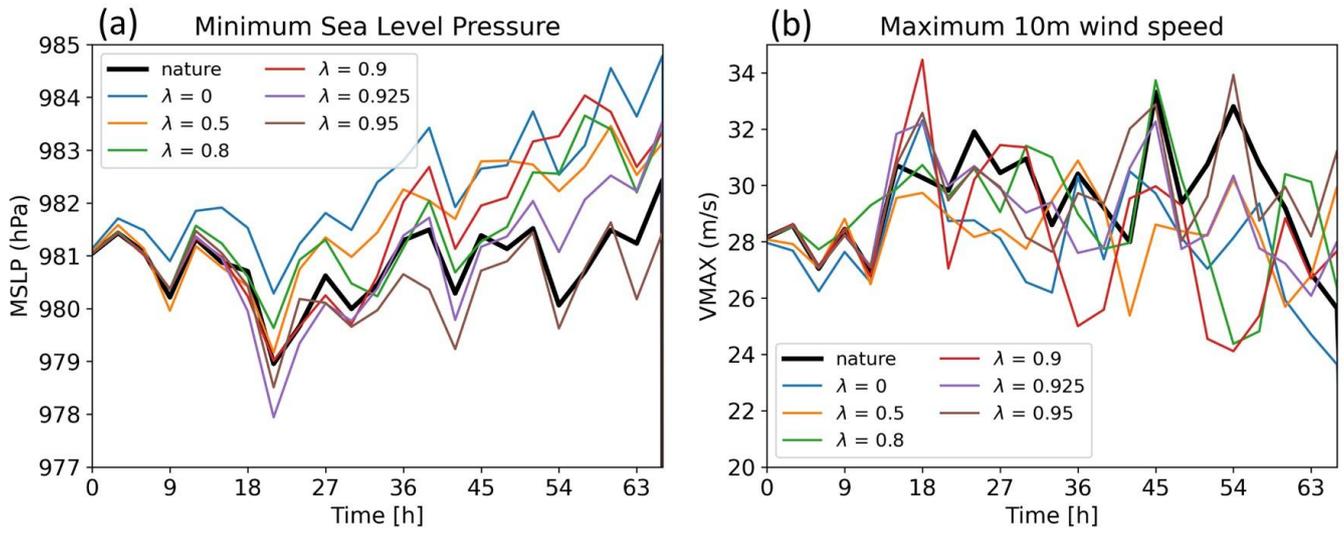

**Figure 3.** (a) Timeseries of minimum sea level pressure in nature run with no control interventions (black) and EnKC experiments with $\lambda$ of 0.0 (blue), 0.5 (orange), 0.8 (green), 0.9 (red), 0.925 (purple), and 0.95 (brown). (b) same as (a) but for maximum 10m wind speed.

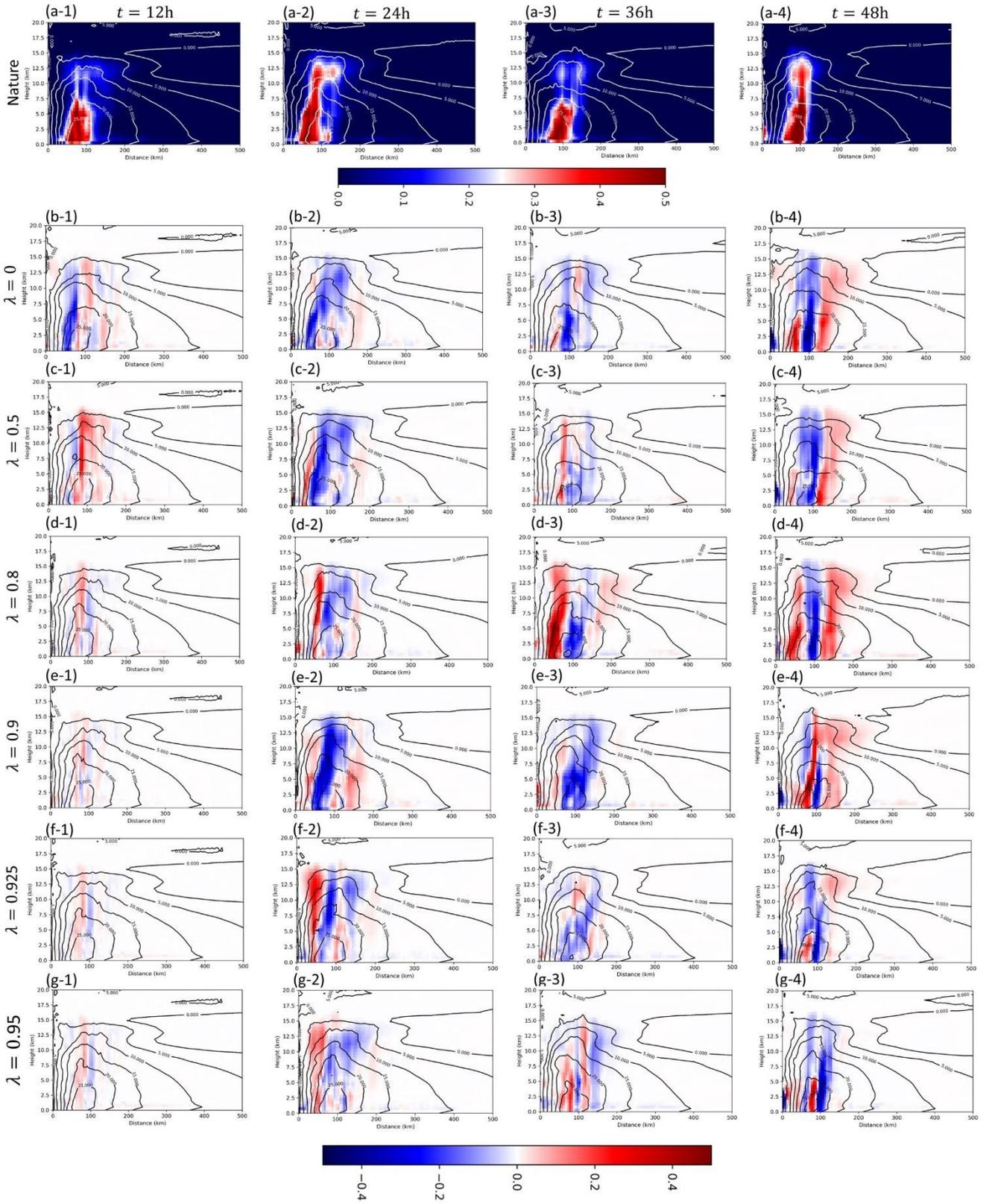

**Figure 4.** (a) Azimuthal average of specific hydrometeors [g/kg] (shades) and tangential wind (contours) in nature run. Horizontal and vertical axes show the distance from the domain center and height, respectively. (b-g) Same as (a), but for the differences of specific hydrometeors between nature and EnKC experiments with $\lambda$ of (b) 0.0, (c) 0.5, (d) 0.8, (e) 0.9, (f) 0.925, and (g) 0.95. Contours show tangential wind of each experiment.

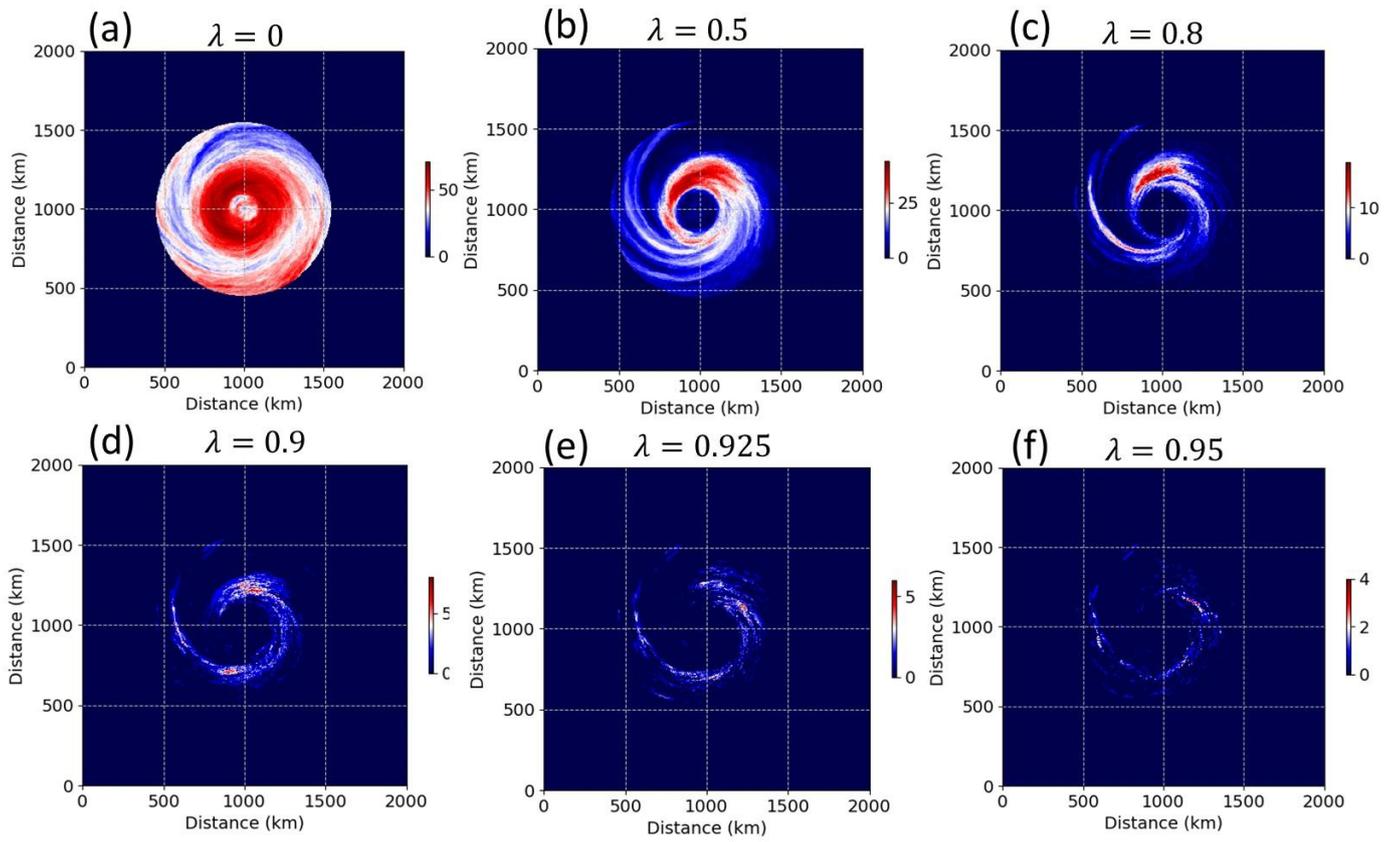

**Figure S1.** The total number of interventions by EnKC with $\lambda$ of (a) 0.0, (b) 0.5, (c) 0.8, (d) 0.9, (e) 0.925, and (f) 0.95. Note that the color scales are different in the different panels.

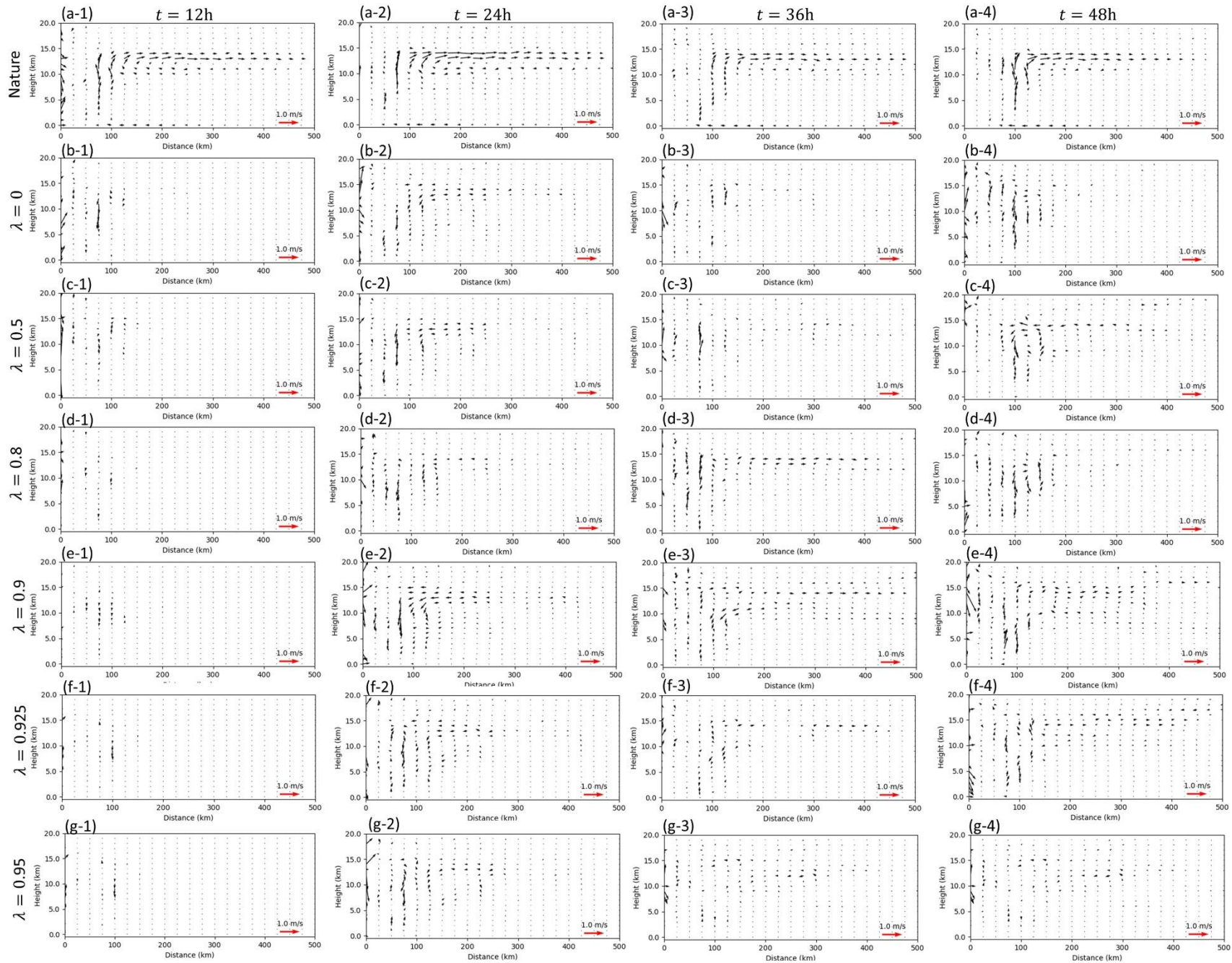

**Figure S2.** (a) Azimuthal average of wind [m/s] in nature run. Horizontal and vertical axes show the distance from the domain center and height, respectively. (b-g) Same as (a), but for the differences between nature and EnKC experiments with $\lambda$ of (b) 0.0, (c) 0.5, (d) 0.8, (e) 0.9, (f) 0.925, and (g) 0.95.